**Laser-Scrawled Random Plasmonic Metasurface in Nanoseconds for Physical Unclonable Functions**

*Haining Xu[1], Yang Zhang[1],\* Shenqi Yang[1], Zhiwei Yuan[1], Jiahui Jin[1], Kaili Kuang[1], Mingze Liu[1], Qiao Wang[1], Yannan Tan[2], Zhenguo Jing[1], Changyu Shen[3], Yurui Fang[1], Wei Peng[1],\**

[1]School of Physics, Dalian University of Technology, 116024, Dalian, China
[2]Dalian Institute of Chemical Physics, Chinese Academy of Sciences, Dalian 116023, China
[3]College of Optical and Electronic Technology, China Jiliang University, Hangzhou, Zhejiang, 310018, China

E-mail: yangzhang@dlut.edu.cn  Yang Zhang
        wpeng@dlut.edu.cn  Wei Peng

Funding: The authors would like to acknowledge the financial supports from the National Natural Science Foundation of China (NSFC: 62275038, 6257034622, 12574330), National Key Research and Development Program of China (Grant No. 2023YFB3209500), the Fundamental Research Funds for the Central Universities (Grant No. DUT24Y152/DUT25YG110/DUT24BK060), the Natural Science Foundation of Liaoning Province (Grant No. 2023JH2/101700156), Dalian Municipal Health and Wellness Guidance Plan Project (2024ZDJH01PT017).

**Abstract:**

Randomness in optical systems emerges as a powerful resource for generating complex, non-deterministic light-matter interactions. In particular, random plasmonic metasurfaces harness nanoscale disorder to produce unique and irreproducible optical responses, positioning them as an ideal platform for physical unclonable function in secure optical authentication. However, realizing such random metasurfaces in a rapid, scalable, and chemical-free manner for optical PUFs remains challenging. Here, we introduce a nanosecond pulsed laser scribing method for one-step fabrication of a robust random plasmonic metasurface physical unclonable function. By delivering spatially localized, ultrafast energy bursts, this technique harnesses naturally occurring instability to generate stochastic plasmonic nanostructures in nanoseconds. The unique plasmonic metasurfaces are effectively transformed into a macroscopic, non-replicable optical fingerprint via morphology-dependent resonance at the nanoscale, enabling low-cost and fast readout. Leveraging the wavelength-selective plasmonic response, we present a multidimensional multiplexing strategy that expands the challenge response pairs space and encoding capacity by 5-fold via topography and RGB multiplexing. The resulting plasmonic keys exhibit good bit uniformity (average: 0.500), high uniqueness (inter-Hamming distance: 0.499), and large capacity (~$2^{8000}$ bits per PUF), with strong environmental stability and resistance to reverse nanofabrication. This work demonstrates how fast laser induced stochasticity can be rationally harnessed and engineered for optical PUFs, opening pathways toward disorder-enabled photonic devices.

# 1. Introduction

Counterfeiting incurs global economic losses exceeding trillion dollars annually and poses a significant threat to public health, national security, and technological integrity[1]. To address these challenges, physical unclonable functions (PUFs)[2] have emerged as a vital platform for secure authentication, leveraging intrinsic stochasticity and uniqueness to produce identifiers that are theoretically impossible to predict and replicate. Since *Pappu et.al.* first introduced the concept of physical one-way function[3] using coherent multiple scattering, numerous implementations have validated the PUF paradigm as an effective route toward unbreakable hardware-level encryption. Among various PUF modalities, optical PUFs[4–6] have garnered particular interest due to their high entropy[7], parallel-readout, and compatibility with multidimensional encoding capability. Notably, the challenge response pairs (CRPs) can be significantly expanded by multiplexing the intrinsic degree of freedom (DoF) of light such as wavelength[8] and polarization[9], to act as strong optical PUFs. To date, optical PUFs can be broadly classified into two categories. The first directly captures random structural features[5] at the micro- or nanoscale, such as wrinkling[9], nanotexture[10,11], or artificial fingerprints[12], stochastic dewetting patterns[13,14], and randomly located micro/nanoparticles[15–18]—to form unique optical patterns. The second exploited multidimensional stochastic optical responses[19–23], including random stimulated luminescence, surface-enhanced Raman scattering (SERS) patterns[24], photoluminescence (PL) patterns [19,20,25,26], and stochastic scattered pattern[3,8,27]. Despite both strategies demonstrating strong unclonability, they always require sophisticated and time-consuming fabrication processes, including chemical synthesis, multi-step fabrication, templated assembly. These practical constraints limit their scalability, structural complexity, environmental stability, and fast readout, leaving a critical need for a PUF platform that is simultaneously truly random, robust, and compatible with scalable, low-cost manufacturing[28].

Random metasurfaces have recently emerged as a powerful flat photonic platform that transforms subwavelength structural randomness into functional light manipulation[29]. Unlike deterministic metasurfaces designed through rigorous phase engineering, random metasurfaces utilize disorder meta-atom arrangements (both in size and location) to transform sub-wavelength stochasticity into versatile functionalities, such as structural color[30], strong near-field enhancement[31], and energy harvesting[32], just to name a few. In this context, harnessing such non-deterministic and unique optical randomness in robust metasurfaces aligns with the requirements of optical PUFs and thus opens new pathways toward secure optical authentication. Nevertheless, existing fabrication routes for random metasurfaces often rely on lithography[33], masked deposition[8], template-assisted growth, or stochastic assembly[34], which remain costly, slow, or limited in structural complexity. To this end, a fast, single-step, and scalable method for generating random metasurfaces capable of serving as robust optical PUFs remains elusive.

Nature already provides the preeminent model of efficiency toward achieving such unique and complex structure through instability, for instance, a discontinuous cylindrical stream of water from a faucet kick. Harnessing such intrinsic randomness represents an effective strategy for generating random structure even with a deterministic design. In this context, nanosecond pulsed laser ablation (PLA) offers a unique mechanism to trigger, freeze, and record the naturally occurring instability at the nanoscale[35]. Unlike the femtosecond laser, which minimizes thermal effects to achieve high-precision deterministic patterning, the nanosecond laser-matter interaction leads to ultrahigh temperatures with ultrafast heating and cooling rates, enabling extreme nonequilibrium conditions that show the possibility for direct fabrication of random metasurfaces PUF. While extensive efforts have focused on employing fast[36] and ultrafast laser[37] for fabricating ordered metasurfaces[38], the intrinsic stochasticity of laser–matter interactions remains underexplored as a functional advantage.

Here, we draw inspiration from naturally occurring instability for ultrafast fabrication of random plasmonic metasurfaces for optical unclonable functions. We introduce an ultrafast, single-step, and scalable top-down approach for fabricating optical PUF via direct nanosecond pulsed laser scribing. By directly freezing these nanoscale instabilities into the substrate, our method translates inherent randomness at the nanoscale into a macroscopic, non-replicable optical fingerprint with multichannel unclonability. Due to the wavelength-selective response optical feature, the red, green, and blue (RGB) channels of the PUF pattern are irrelevant and can be employed as three additional different spatial intensity information, enabling the expansion of CRPs and multi-channel encryption without the need for additional physical DoF or complex implementation. In conjunction with scalability to large-scale manufacturing, our cost-effective random metasurfaces represent a promising approach for bridging the gap between laboratory-scale demonstrations of optical PUFs and practical, widely deployable authentication technologies.

## 2. Results

### 2.1. Nanosecond pulsed laser scribing of random plamonic metasurface PUFs

**Figure 1a** schematically illustrates the nanosecond laser scribing approach for generating random plasmonic metasurface PUFs. We choose the material of the random metasurface to be silver, one of the most commonly used noble metals due to its strong plasmonic response. First, a silver nanofilm was deposited on a SiO$_2$ substrate via magnetron sputtering. To ensure convenient demonstration and consistent PUF keys readout, a shadow mask was employed to roughly define the active Ag nanofilm region as 1 mm × 1 mm (Figure 1b and Figure S1, Supporting Information). The defined region was then subjected to tightly focused rectangular shape nanosecond pulsed laser irradiation, triggering localized ablation on a nanosecond timescale and enabling rapid fabrication in seconds without any post-treatment.

The core of our random plasmonic metasurface design is to exploit and freeze the laser-triggered Rayleigh–Plateau (RP) instability[39] within nanoseconds, transforming transient molten disorder into permanent structural randomness. Upon nanosecond laser irradiation (Figure 1a), photon absorption rapidly excites conduction electrons (photon-electron coupling, ~5 fs), followed by rapid electron-phonon energy transfer (electron-phonon coupling) and lattice heating.[40] The two-temperature model (TTM) simulation (Figure 2a, see more details in Supplementary Note 1) reveals that continuous photon absorption drives the lattice temperature to ~1700 K, far exceeding the melting point (~1234 K) of silver, leading to a transient molten layer prone to fragmentation driven by Rayleigh–Plateau instability. Sustained laser energy further destabilizes the molten layer, inducing rapid capillary fragmentation under transient superheated conditions. Subsequently, rapid heat exchange (cooling rate ~$10^7$ K s$^{-1}$) drives spontaneous recrystallization of the transient molten silver pools. During dewetting, the liquid droplets minimize their surface energy at the metal–air–substrate interfaces, adopting equilibrium hemispherical morphologies (Figure 1d-f) governed by capillary forces. Because the lifetime of the molten droplets is shorter than the hydrodynamic relaxation, nanoscale instabilities are kinetically trapped, yielding a densely packed, randomly disordered array of nano-hemispherical structures (Figure 1d-e) and sub-micron voids along each scribed line (Figure S3, Supporting Information). High-resolution transmission electron microscopy (HR-TEM) images (Figure2b and FigureS4, Supporting Information) reveal the stacking faults in silver nanoparticles, corroborating the ultrafast cooling and solidification that preserve nanoscale instabilities as permanent structural features[41].

These nanoscale features exhibit substantial spatial variability in particle size, distribution, and density across the scribed regions, effectively preserving the transient instability induced during the PLA process. Intriguingly, the nano-hemispheres exhibit a hierarchically trimodal size distribution (~sub-10 nm, 40 nm, and 250 nm), with the smallest particles often clustering around the larger ones (Figure 1e-f and Figure S5, Supporting Information). These observations suggest that the SEM images represent a frozen snapshot of the dynamic laser ablation and dewetting evolution (see more details in Supplementary Note 2).

Disorder is intrinsically ensured during the process by nanosecond laser-triggered RP instability. For a quantitative illustration, we perform a two-dimensional Fast Fourier Transform (2D-FFT) based on the position of the Ag nanoparticles. A circle-shaped distribution of the squared Fourier components is observed in Figure 2g and Figure 2h, confirming the disordered arrangement of Ag nanoparticles. To further evaluate the nanoscale uniqueness of each random plasmonic metasurface, the Scale-Invariant Feature Transform (SIFT) algorithm[42], which detects and describes local features invariant to scale, rotation, and illumination changes, was employed to compare across different dark field (DF) imaging[18]. As shown in Figure 2c, feature matches between identical PUFs are highlighted in green, while false matches appear in blue. The inlier ratio quantifies the matching fidelity—the proportion of geometrically validated matches among all feature pairs. The resulting correlation matrix (Figure 2d) visualizes the one-to-one correspondence among identical PUFs, validating nanoscale uniqueness. Additionally, random sub-3 nm inter-particle spacing (Figure 1d and Figure S6, Supporting Information) emerges randomly throughout the nanostructure, introducing additional entropy via random and complex near-field coupling.

At the microscale, the tightly focused laser profile creates unpredictable valley-like scribed patterns (Figure 1c and Figure S7, Supporting Information). The stochastic scribing process introduces spatial heterogeneity in number, length, orientation, and scribed fluence, leading to singly or multiply scribed regions. Such fluctuations in laser-matter interaction generate irregular surface topographies, where the underlying nanoscale hills are embedded in the microscale valleys. The stochastic Ag nanoparticles exhibit morphology-dependent resonant absorption, translating underlying nanoscale randomness into macroscopic optical textures. The resulting multi-scale disordered structure serves as an ideal platform for optical PUFs, whose uniqueness and complexity are fundamentally resistant to replication, even with state-of-the-art nanofabrication techniques. Consequently, vivid and colorful PUF patterns emerge instantly as the random plasmonic metasurface is written in a single step (Figure 2b-ii), drastically reducing fabrication cost and time. The distinctive, randomly colored furture enables the digitization of binary codes, endowing intrinsic anti-counterfeiting and anti-duplication functionalities.

## 2.2. Random plasmonic metasurface PUF

The unpredictable and irreproducible metasurface is directly transformed into a microscale optical fingerprint via plasmonic structural coloration, enabled by localized and gap plasmon resonances in densely packed nanostructures. This stochastic color information can be discriminated by a Bayer filter constructed in a commercial imaging sensor, resulting in three RGB channels that provide three different spatial intensity information within a single plasmonic PUF pattern. As shown in Figure 2i, the optical features of the laser scribed pattern are found difference for both spatial intensity and localized structural details (marked with white circle in Figure 2i) under red, green, and blue filtered light illumination, indicating wavelength-dependent light–matter interactions that reflect the intrinsic nanostructure randomness.

The DF optical microscope was employed to reveal the random plasmonic structure color generated by the PUF pattern (Figure 2j). Unlike the typical bright spot associated with an isolated particle[43], the laser-scribed region exhibited a galaxy-like multicolored scattering pattern, which is excited from the highly packed nanoparticle structure. This characteristic significantly increases our optical PUF's structural complexity and its resistance to cloning during reverse nanofabrication. Due to the stochastic scribing, distinct color hues, dominated by blue or orange scattering, are spatially segregated along the laser lines. Moreover, the color also varies across the scattering region, particularly near the line edge, where local nanoparticle morphology fluctuates the most (Figure 2k). The DF scattering spectrum of four representative points with different scattering colors (marked on Figure 2i) exhibited clearly different scattering peaks (Figure 2l), strongly indicating the random and heterogeneous nanostructure profile of the corresponding region. The representative SEM images (Figure 2e–f) under a microscale field of view captured at the regions along the laser-scribed line elucidated pronounced differences in nanoparticle morphology, which is the underlying core of the optical fringprints.

To elucidate the underlying plasmonic coupling mechanisms for the random metasurface, we numerically employed a semi-analytical model to reproduce the geometry profile of the laser-scribed nanostructures (see more details in Supplementary Note 3). Specifically, a random disordered Ag nano-hemisphere array is built in which both particle size and spatial position are randomized. Typically, the RGB components could relate to the spectrum response of three spectral ranges (Red: 600-700 nm, Green: 500-600 nm, Blue: 360-500 nm) respectively, and we select three representative wavelength positions for illustration ($\lambda_R$= 700 nm, $\lambda_G$= 550 nm, $\lambda_B$= 450 nm). The dip in simulated transmission curves (namely, absorption) around 450 nm mainly results from the dipolar resonance of isolated nanoparticles (Figure 2m), where the electric field is strongly amplified within individual Ag nano-hemispherical profiles (Figure 2n-i). In contrast, this size-related LSPR resonance is tuned off at 700 nm. Nevertheless, the electric profiles (Figure 2b-ii) show a tightly confined field of gap plasmon modes in the narrow spacing between particles at both in- and off-resonance wavelengths. This off-resonance feature is attributed to lateral hybridization of near-spacing nanoparticles (namely, oligomers). This is consistent with experimental observations (Figure 1d), where interparticle spacing below 3 nm is commonly found and constructs oligomers[18]. For the green channel (550 nm), the electric field profile exhibits both isolated plasmonic resonance and gap mode, compared to the R and B wavelengths. Yet, a mild electric field amplification is observed, resulting in a smaller absorption (a larger transmission) compared to the well-exited LSPR wavelength. It is expected that other nanostructure configurations will yield a rich spectrum of structural colors. As such, this wavelength selective optical feature that depends on nanoscale structure randomness enables the RGB components of the same pixel to arise from nanomorphology-dependent optical resonances, and are therefore not trivially correlated [8]. Additionally, regions that remain unirradiated preserve the continuity of the Ag nanofilm and exhibit a characteristic blue-violet metallic sheen under bright-field illumination (Figure S1, Supporting Information). This feature broadens the effective color gamut of the plasmonic PUF pattern, particularly enhancing the representation within the blue channel.

**2.3. PUF key generation and security performance**

The fundamental requirement for any PUF fabrication strategy is the ability to generate unique patterns with a deterministic design. To quantify the similarity between macroscopic plasmonic PUF patterns, we applied feature-based matching to evaluate their pairwise correspondence. As shown in Figure 3a, identical PUF patterns yielded a high inlier ratio of 0.793 (Figure 3a) even under deliberate perturbations and repeated imaging (Figure S9, Supporting Information). Conversely, distinct plasmonic PUF patterns exhibited a significantly lower inlier ratio of 0.162, confirming their intrinsic dissimilarity. A threshold of 0.3 for the inlier ratio and a minimum of 40 matches was set to distinguish between identical and different patterns. This criterion was applied to generate a uniqueness matrix (Figure 3b), where only diagonal elements exceeded the threshold, affirming the uniqueness of our nanosecond laser fabrication.

The inherent randomness at both the nanoscale and microscale endows our random plasmonic metasurface PUF pattern with multimode multiplexing capability, enabling multi-level key generation under a single optical challenge and straightforward bright-field optical imaging. Based on the acquired optical spatial-intensity information (grayscale for representation), two distinct types of cryptographic keys were first extracted to support multi-level security applications. The first, topography-based key leverages the random micro-scale valley-like surface features (Figure 3c, upper row) of the plasmonic PUF pattern. The captured intensity distribution effectively reflects the valley topography, allowing rapid image-based reconstruction without the need for time-consuming confocal microscopy. To enhance the topography extraction and key generation, an adaptive threshold is applied to extract a binary

'0-1' binary map (40×40 pixels), where the threshold for each pixel is dynamically determined based on its local neighborhood value (Figure S10, Supporting Information). The Hamming distance (HD, see more details in Supplementary Note 4), which measures the bit-state difference of the same pixel position across the binary map, was used to test the key performance. The resulting similarity matrix (HD of different topography keys, Figure 3f) shows distinctiveness and even with noise perturbation (Figure S11, Supporting Information). The extracted morphology-based key enables rapid, pre-verification and provides an effective initial barrier against unauthorized replication.

The topography-based key omits the fine intensity variations introduced by nanoscale random plasmonic absorption. For high-level secure keys, the complete spatial intensity profile of the plasmonic PUF pattern was processed through a cryptographic one-way hash function (Figure 3c and Supplementary Note 4), producing a robust binary code resistant to modeling and reverse engineering. This one-way hashing ensures that the generated key cannot be reconstructed from the output, providing strong protection against key inference attacks. To further improve stability and reduce susceptibility to environmental perturbations, the PUF images were transformed via the discrete 2D Fourier transform into the Fourier domain, with only the low-frequency components (first 40 × 40 pixels) retained to generate the final 1,600-bit key. Key quality was evaluated using standard performance metrics, including bit uniformity, inter- and intra-Hamming distance. As shown in Figure 3e, the bit uniformity approaches the ideal value of 0.5, indicating that the generated keys are both balanced and unpredictable, exhibiting characteristics of a true random number generator. Inter-Hamming distances (Figure 3d) exhibited near-ideal separation with a mean value of 0.499 and standard deviation of 0.025 by fitted to a Gaussian distribution, confirming uniqueness across the key space. Intra-Hamming distances remained low under repeated measurements loaded with noise (mean value of 0.067 and standard deviation of 0.012), demonstrating key reproducibility and robustness for repeated authorizations. For practical authentication, re-capturing the PUF pattern may introduce a spatial misalignment due to device positioning variations. To ensure reliable key extraction, the PUF image can be aligned in the Fourier domain using phase correlation (Supplementary Note 5)[23] before the authentication:

$$F(u,v) = G(u,v) \exp\left[2\pi i \left(\frac{udx}{M} + \frac{vdy}{N}\right)\right] \tag{1}$$

where $F(u,v)$ represents the Fourier transform of the raw plasmonic PUF pattern, and $G(u,v)$ represents the Fourier of the re-captured plasmonic PUF image during authentication. The exponential term $\exp\left[2\pi i \left(\frac{udx}{M} + \frac{vdy}{N}\right)\right]$ encodes the spatial shift as a linear phase change in the frequency domain. By inverting this phase shift (Figure S12, Supporting Information), the original alignment is restored, ensuring consistent PUF key extraction regardless of capture position deviations.

Also, the false positive rate (FPR) and false rejection rate (FRR) are evaluated from the Gaussian fitting of intra- and inter-Hamming distance to assess the probability of wrong authorizations. The result (Figure S13, Supporting Information) shows a negligible number of ~$1\times10^{-126}$ for the cross point, which means nearly successful authorizations are obtained by setting the separation threshold of 0.194 (the cross point of the intra- and inter-Gaussian fitting). Moreover, the entropy (the number of independent bits) of our PUF, the effective number of independent bits (ENIB), gives an aspect for measuring the randomness and independence of bits for a plasmonic PUF, as calculated as follows:

$$N = \frac{\mu_{inter} \times (1 - \mu_{inter})}{\sigma_{inter}^2} \qquad (2)$$

where μ is the mean value and σ is the standard deviation of distribution for inter- and intra Hanmming distance. The entropy is a crucial parameter for validating the security level of PUF keys, where a large entropy means more security against brute force attacks. For only a gray scale image, ENIB are extracted of 1275 and subsequently result in an encoding capacity of $2^{1275}$.

### 2.4. RGB multiplexing of plasmonic PUFs

Unlike conventional PUFs that encode multidimensional physical information, which typically record one spatial intensity information under one single challenge[44], our random plasmonic metasurface PUF inherently carries three RGB channel information due to the stochastic plasmonic coloration. We for the first time introduce a simple and applicable RGB multiplexing strategy to expand the CRP's space by a single bright field optical image under a single optical challenge (Figure 4a).

To validate the feasibility of RGB multiplexing, individual grayscale images corresponding to the R, G, and B channels were extracted and analyzed independently. As shown in Figure 4b, cross-channel comparisons using the SIFT algorithm revealed low inlier ratios and insufficient matching lines between different color channels of the same plasmonic PUF pattern (inlier ratio < 0.3, match lines < 40), indicating minimal correlation and statistical independence among the channels. This orthogonality is further illustrated in the cross-channel similarity matrix (Figure 4c and Figure S14, Supporting Information), where only self-channel comparisons exhibit high matching fidelity. Notably, the spatial intensity information captured in one color channel cannot be inferred or reconstructed from the others, owing to their independent optical responses on the same pixel, which ensures that each RGB channel contributes unique and secure.

Before hash key generation, we introduce a "color mask" (Supplementary Note 6) to robustly convert the colorful pattern as "true RGB" spatial information, following the principle of multi-color fluorescent PUFs[44,45] or wavelength multiplexing encoding[8]. By setting a consistent color threshold across all the plasmonic PUF patterns, the wavelength-selective response optical fingerprint was transformed to three exclusive RGB spatial intensity information (Figure 4a and Figure S15). Subsequently, three additional PUF keys (also processed as 40×40 bits for demonstration) could be extracted by processing the masked RGB channel via a standard Hash algorithm. These color-specific keys demonstrated the same quality as the grayscale-derived key, exhibiting near-ideal bit uniformity (Figure 4d) and inter/intra-Hamming distance distributions (Figure 4e) indicative of high randomness and uniqueness. The similarity matrix (Figure 4f) suggests the distinctiveness of RGB channels across all multiplexed plasmonic PUF patterns. Notably, this approach significantly enhances the encoding capacity. Whereas prior strategies have relied on multiplexing multidimensional physical modalities (e.g., Raman scattering or multi-color fluorescence, or with lifetimes multiplexing) to increase CRP's space, our strategy achieves comparable expansion through channel multiplexing alone, thanks to the intrinsic wavelength-selective optical feature and the practical "color mask" strategy. Alternatively, by compromising readout portability for enhanced security, the complex scattering patterns in DF images can serve as a high-entropy PUF source (Figure S16, Supporting Information).

By concatenating the two grayscale-derived keys with the three RGB-channel keys, we generated a composite 8000-bit binary key. This represents a total encoding capacity of $\sim 2^{8000}$ bits per plasmonic PUF pattern, achieved solely through spectral channel decomposition without multiplexing physical parameters (Figure 4g). This capacity can be dynamically tuned by selecting frequency bands in the Fourier domain (Figure S17, Supporting Information).

## 2.5. Cryptographic Viability and Demonstrations Enabled by Multichannel PUF Keys

Beyond authentication, we further evaluate the cryptographic suitability of the expanded plasmonic PUF-derived keys. Standard NIST SP 800-22 randomness tests were performed on 8000-bit key streams extracted from our plasmonic metasurfaces. As shown in Figure 5a, all statistical tests were passed, confirming that the PUF-generated keys exhibit the statistical properties of a true random number generator.

To demonstrate a practical cryptographic application (Figure 5b), we employed the two independent keys for symmetric encryption. These two keys (denoted A and B) were then combined via a bitwise exclusive-OR (XOR) operation (C=A⊕B) to produce a shared public key (C), which was stored in the data cloud. To encrypt a message, a 40×40-pixel binary cartoon image was XOR-encrypted using key A, and key B for decrypted by XOR-mixing the public key-C (A=C⊕B). Although both keys derive from the same PUF image, their distinct encoding mechanisms guarantee independence and unpredictability, eliminating the need for secret key sharing as the public composite key (A⊕B) is also inherently unique and irreproducible.

Leveraging the stereoscopic encoding capability of our plasmonic PUF, we further demonstrate using a single random plasmonic metasurface for symmetric encryption a colorful image (Figure S18, Supporting Information). A colorful cartoon image was encrypted by applying XOR operations between each RGB channel and the corresponding PUF-derived key. The resulting ciphertext, featured as a random, colorful QR code, was sent to the receiver. Only with the same plasmonic PUF key could the ciphertext be decoded to the original image. With loosed of any channel information of the plasmonic PUF key, the deciphered text could not reproduce the colorful cartoon chicken and remains a messy code.

## 2.6. Robust random plasmonic metasurfaces and versatility of nanosecond laser fabrication

For practical readout and authentication, our plasmonic PUF patterns can be readily imaged using a portable smartphone-based microscope. A representative optical image captured by the smartphone camera is shown in Figure S19a, Supporting Information. Despite inherent differences in resolution and image processing algorithms, the plasmonic PUF image and extracted keys still exhibit a high inlier ratio, confirming reliable matching performance. These results demonstrate the potential for real-world deployment using standardized devices and authentication protocols in mobile and edge-security settings. We further illustrate a conceptual lensless CMOS-based platform (Figure S19b, Supporting Information), suggesting the potential for ultra-compact, low-cost authentication systems.

Stability is a critical requirement for reliable PUF tag deployment in real-world applications. We systematically evaluated our random plasmonic metasurfaces' reproducibility across a range of environmental perturbations (Figure 5c). The PUF chips were sequentially subjected to temperature variation (-20 to 70 °C), ultrasonic vibration (to test the adhesion of the metal nanostructure to the substrate), and long-term sunlight exposure (5 months). In all cases, the resulting keys exhibited both low inliner ratio and HD, confirming robust key stability and minimal bit flipping under mild physical disturbances (Figure 5d). Re-authentication after five months remains high inliner ratio, suggesting high stability of the random plasmonic metasurface PUF. Such reproducibility is critical for enabling repeated authentication in long-term and field-deployed applications. For real-world implementation, an optically transparent protective layer can be applied to the surface (Figure S19c, Supporting Information), further enhancing durability without compromising readability.

As a versatile fabrication technique, nanosecond pulsed laser scribing is demonstrated to be fully compatible with flexible polymer substrates such as polydimethylsiloxane (PDMS). Unlike conventional PUF fabrication methods, such as high-temperature annealing or multistep wet-chemical synthesis, which are incompatible with low–thermal–tolerance substrates, our approach enables direct, one-step patterning under ambient conditions (Figure 5c). These flexible, elastomeric polymer films conformally adhere to various surfaces, including curved and irregular geometries such as pharmaceutical packaging or official documents (Figure S19c, Supporting Information) for secure anti-counterfeiting.

To demonstrate the scalability of our nanosecond laser scribing method, a colorful school logo (25mm×25mm) was readily fabricated with the help of a customized shadow mask, as shown in Figure 5f. Beyond the microscopic digital keys, the macroscopic plasmonic PUF itself serves as an instant, naked-eye verifiable security label (Figure 5g). Due to the complementary nature of plasmonic resonance, the Ag-based logo displays a translucent yellow hue in transmission (subtractive color) but flips to a vivid blue scattering color under oblique scattering in dark background observation. This striking chromatic contrast arises from the strong LSPR scattering of silver nanoparticles, which predominantly scatter in the blue-violet spectrum. Considering its simplicity, speed, damage-free nature, and robust readout (see more information in Supplementary Note 6), our nanosecond laser scribing strategy offers a competitive and versatile route for optical PUF generation, outperforming the state-of-the-art approaches (Figure 5h) such as wet-chemical synthesis, stochastic deposition, or LIPSS.

## 3. Conclusions

In summary, we have demonstrated a random plasmonic metasurface by leveraging non-deterministic, irreproducible light–matter interactions for secure and intrinsically unclonable optical authentication. By harnessing the intrinsic spatiotemporal instabilities of nanosecond laser–matter interactions, our approach enables the spontaneous formation of broadband stochastic nanostructures, featuring sub-3 nm gaps—geometries that are fundamentally irreproducible and thus ideally suited for physical unclonability. These nanoscale instabilities are directly frozen into the substrate within nanoseconds, translating inherent randomness into a macroscopically visible, non-replicable optical fingerprint without the need for chemical synthesis, photolithography, sample alignment, or multi-step post-processing.

The resulting optical PUFs exhibit high complexity, distinctive optical fingerprints, and excellent readout accessibility, requiring only low-magnification optical microscopy or even consumer-grade smartphone imaging. Furthermore, the wavelength-selective plasmonic response provides three independent RGB challenge–response channels, offering multichannel unclonability and significantly expanding the encoding capacity without invoking additional physical degrees of freedom or complex interrogation schemes. Compared with existing structure-based or multidimensional optical PUF strategies, the nanosecond laser scribing method uniquely combines intrinsic stochasticity, rapid fabrication, chemical-free processing, and compatibility with large-area production. Notably, considering the material diversity[25,46], our proposed universal nanosecond laser scribing method gives a new degree of freedom for future PUF design (Figure S20, Supporting Information)

Beyond secure optical PUF for secure authentication, the underlying mechanism—freezing ultrafast instability-driven nanostructures—opens new possibilities for constructing disorder-driven photonic platforms. We believe that our random plasmonic metasurface platform represents a substantive advance toward scalable, disorder-engineered nanophotonics.

## 4. Experimental Section

*Fabrication of Structure Color PUF:* The fabrication flow of the random plasmonic metasurface PUF is illustrated in Figure 1a. In a typical procedure, a 10mm×10mm silica was cleaned with acetone, ethanol, and deionized water successively. The cleaned substrate was deposited with a 30 nm Ag film using a Magnetron sputter (Quorum-Q300TD plus). A shadow mask with 1mm×1mm square hole array (3×3 array) fabricated by 3D-printing was used to define the plasmonic PUF region. The laser scribing was processed using a DUV nanosecond laser (193nm, pulse duration:5ns, MLase-500, MLase GmbH) with this fixed parameter (pulse energy: 2mJ, repetition rate: 20 Hz). The laser beam intensity features a Gaussian profile and is focused by a cylindrical lens. The samples were manually scribed around the laser focal point, naturally resulting in a stochastic variation in the scribing power, scribing area, scribed line pattern (scribled line number, length, angle), and whether re-scribed.

*Characterization:* The nanostructure of the random plasmonic metasurface PUF nanostructure was analyzed by scanning electron microscopy (SEM, JSM-7900F JEOL, USA). Transmission electron microscopy (TEM) and high-resolution TEM (HR-TEM, FEI TF30) were employed for detailed structural analysis. Digital photographs of the PUF chips in Figure 5b were captured using a smartphone (iPhone 13, Apple) integrated with a portable macro lens. The topography of the random plasmonic metasurface PUF surfaces at the nano scale was analyzed with AFM (Dimension FastScan, Bluke, USA) in the tapping mode. The microscale 3D topography was obtained by a scanning laser confocal microscope (LSM900, Carl Zeiss Microscopy GmbH).

*Optical measurement setup:* The bright field image of random plasmonic metasurface PUFs patterns was imaged by an inverted microimaging system (CKX53, OLYMPUS Tokyo, Japan) using an objective lens (4×, NA: 0.10, WD: 50mm) and a color charge-coupled device camera (CCD, MTR3CMOS20000KPA, ToupTek Photonics). The raw images have a resolution of 5440×3648 pixels. The central region (2000×2000 pixels) of the raw image was selected for analysis and converted into a grayscale representation. The dark field image and spectrum were recorded on an Olympus BX51 upright microscope (100×, NA: 0.10, WD: 50mm) that consists of a quartz-tungsten-halogen lamp.

*Simulation:* The transmissive spectra and electric field distribution of the simulated samples were calculated using the experimental geometrical parameters extracted from the scanning electron microscopy (SEM) analysis of the samples, using commercial FDTD software (Lumerical FDTD, Lumerical Solutions Inc.). The relative permittivities of silver are taken from the literature. We set the simulation zone to be 1μm × 1 μm to approximately emulate one pixel on the captured image (that is, roughly 500×500nm/pixel). The temporal temperature profile was simulated using COMSOL Multipphysics.

**Acknowledgements**

The authors would like to acknowledge the financial supports from the National Natural Science Foundation of China (NSFC: 62275038, 6257034622, 12574330), National Key Research and Development Program of China (Grant No. 2023YFB3209500), the Fundamental Research Funds for the Central Universities (Grant No. DUT24Y152/DUT25YG110/DUT24BK060), the Natural Science Foundation of Liaoning Province (Grant No. 2023JH2/101700156), Dalian Municipal Health and Wellness Guidance Plan Project (2024ZDJH01PT017).

**Conflict of interest**

The authors declare no competing interests.

**Data Availability Statement**

The Matlab code for plasmonic PUF key generation, matching, and the plasmonic PUF image data are available from the authors upon reasonable request.

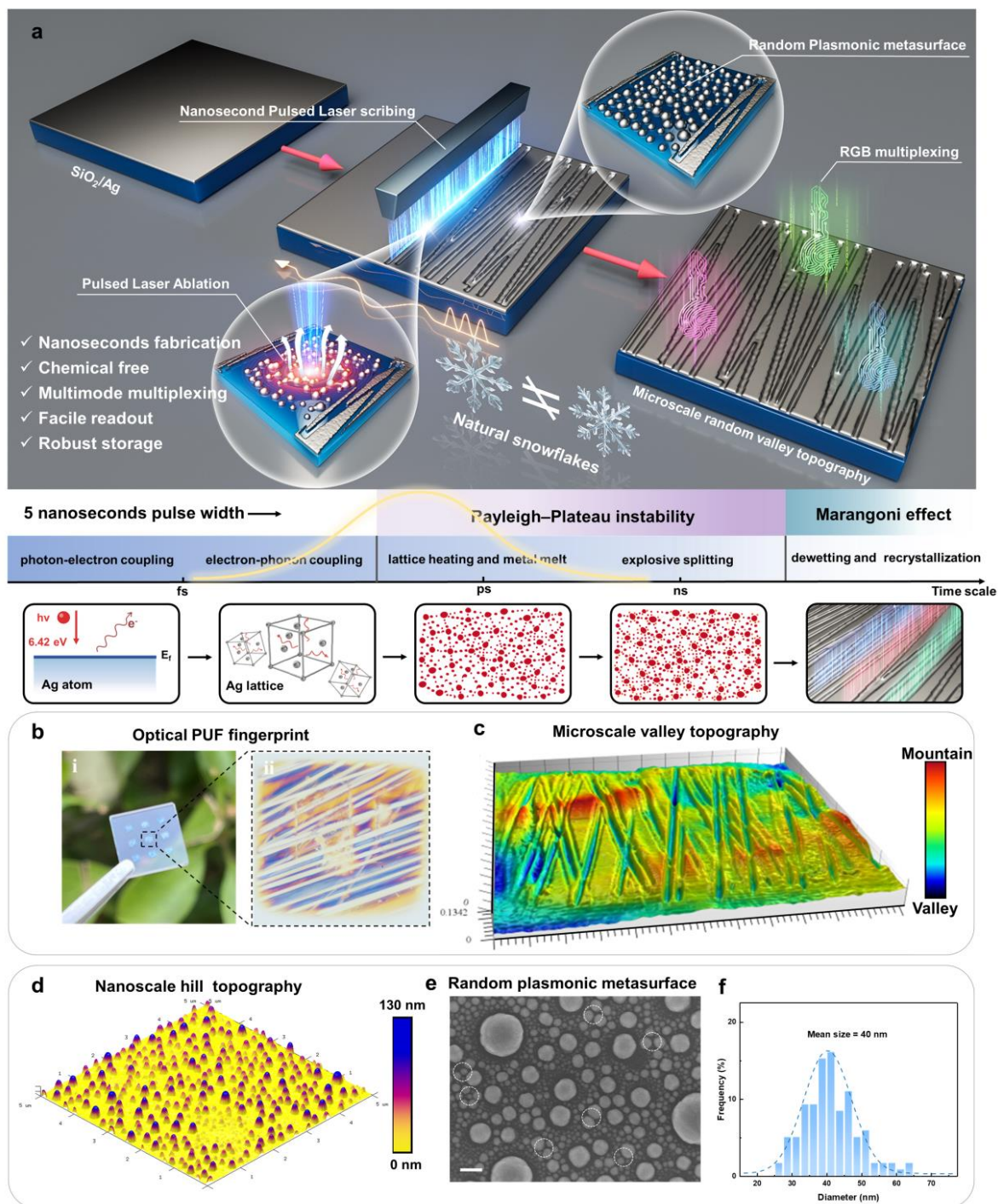

**Figure 1** | Fabrication of random plasmonic metasurface physical unclonable functions. (a) Schematic illustration of a one-step nanosecond laser fabrication process. (b) Photograph of a representative plasmonic PUF chip and its zoom-in optical image, showing unique, unclonable disorder patterns, scale bar, 250μm. (c) Microscale 3D optical profilometry of the random valley topography of a plasmonic PUF using laser confocal microscopy, where mountain–valley fluctuations arise from stochastic laser scribing. (d) Nano-scale 3D morphology inside a laser scribed line using atomic force microscope, revealing densely distributed nanoscopic hemispherical features. (e) SEM image of the random plasmonic metasurface composed of

size-varied Ag nanoparticles. scale bar, 100nm. (f) Size distribution of middle-sized Ag nanoparticles in (e).

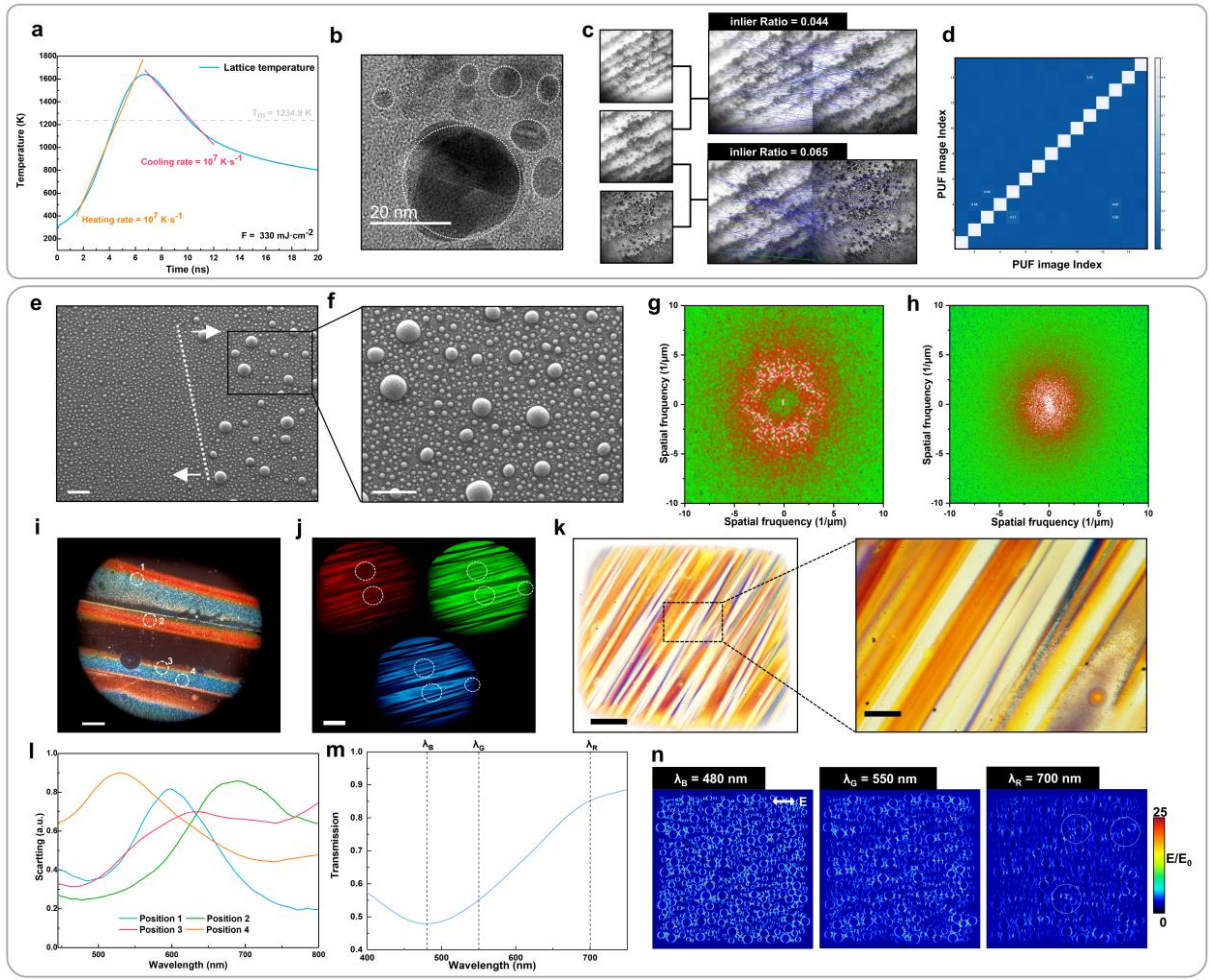

**Figure 2** │ Characterizations of random plasmonic metasurfaces. (a) Two-temperature model simulation of the temperature evolution of nanosecond laser irradiation. (b) TEM image of the nanosecond laser fabricated silver nanoparticles with abundant stacking faults. (c) SIFT-based feature matching between independently captured random plasmonic metasurface DF images, illustrating low inlier ratios. (d) Similarity matrix of different DF images using SIFT-based feature matching, suggesting the nanoscale uniqueness of the random plasmonic metasurfaces. (e) and (f) Representative SEM images of random plasmonic metasurfaces along the laser-scribed lines, which reveal the physical origin of the heterogeneous scattering colors and optical fingerprints observed in plasmonic PUF. Scale bar, 100 nm. (g) and (h) 2D Fourier spectra of corresponding SEM in (e) and (f), showing isotropic and random spatial-frequency distributions corresponding to the stochastic nanoparticle arrangements. (i) DF optical micrograph exhibiting position-dependent scattering colors along the laser scribing lines. (j) RGB-channel–resolved scattering images showing distinct color responses at different positions. (k) Optical

microscopy of extended laser-written stripes, with a magnified view highlighting microscale color fluctuations induced by local nanostructure variations. (l) DF scattering spectrum of four representative points in (i), marked in white circle. m Simulated transmissive spectrum of a random disorder silver nanoparticle array. (n) Optical near-field enhancements of the random plasmonic metasurface at 480, 550, and 700 nm.

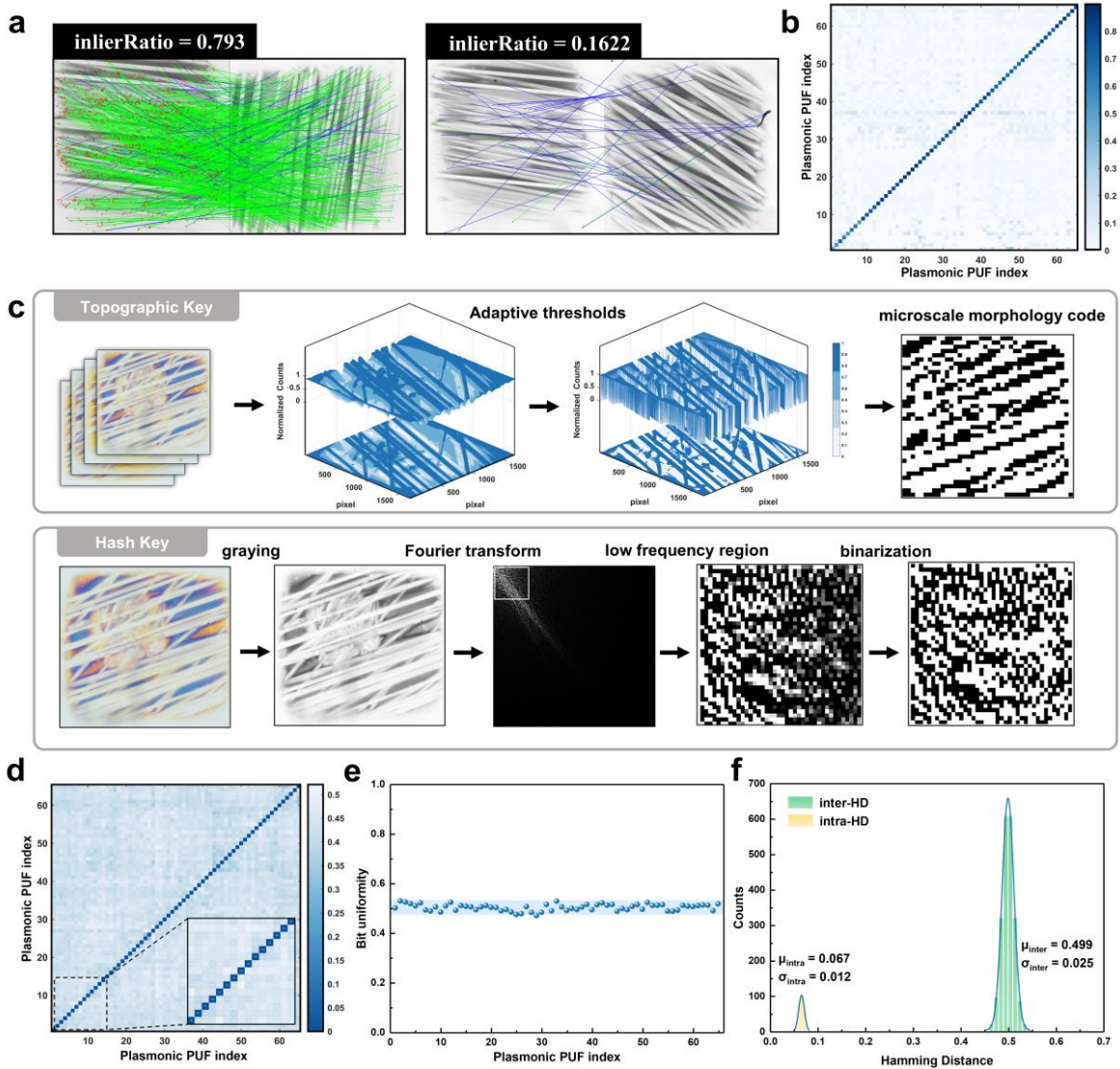

**Figure 3** │ Plasmonic PUF key generation and performance. (a) Feature matching by the SIFT algorithm for the same (left, rotated 90 degrees and loaded with noise) and different (right) PUF pattern. (b) Heatmap of inlier-Ratio for matching identical and different plasmonic PUF patterns with random noise loaded. (c) First panel: Encoding strategy using the graying scale plasmonic PUF pattern, valley topology encoding by adaptive threshold, second panel: high-level plasmonic PUF keys generation using the full spatial intensity information via Hash algorithm. d Similarity matrix of topology-encoded keys under noise perturbations. (e) Bit

uniformity of different plasmonic PUF keys via Hash encoding. (f) inter- and intra-Hamming distance of the same and different plasmonic PUF keys via Hash encoding.

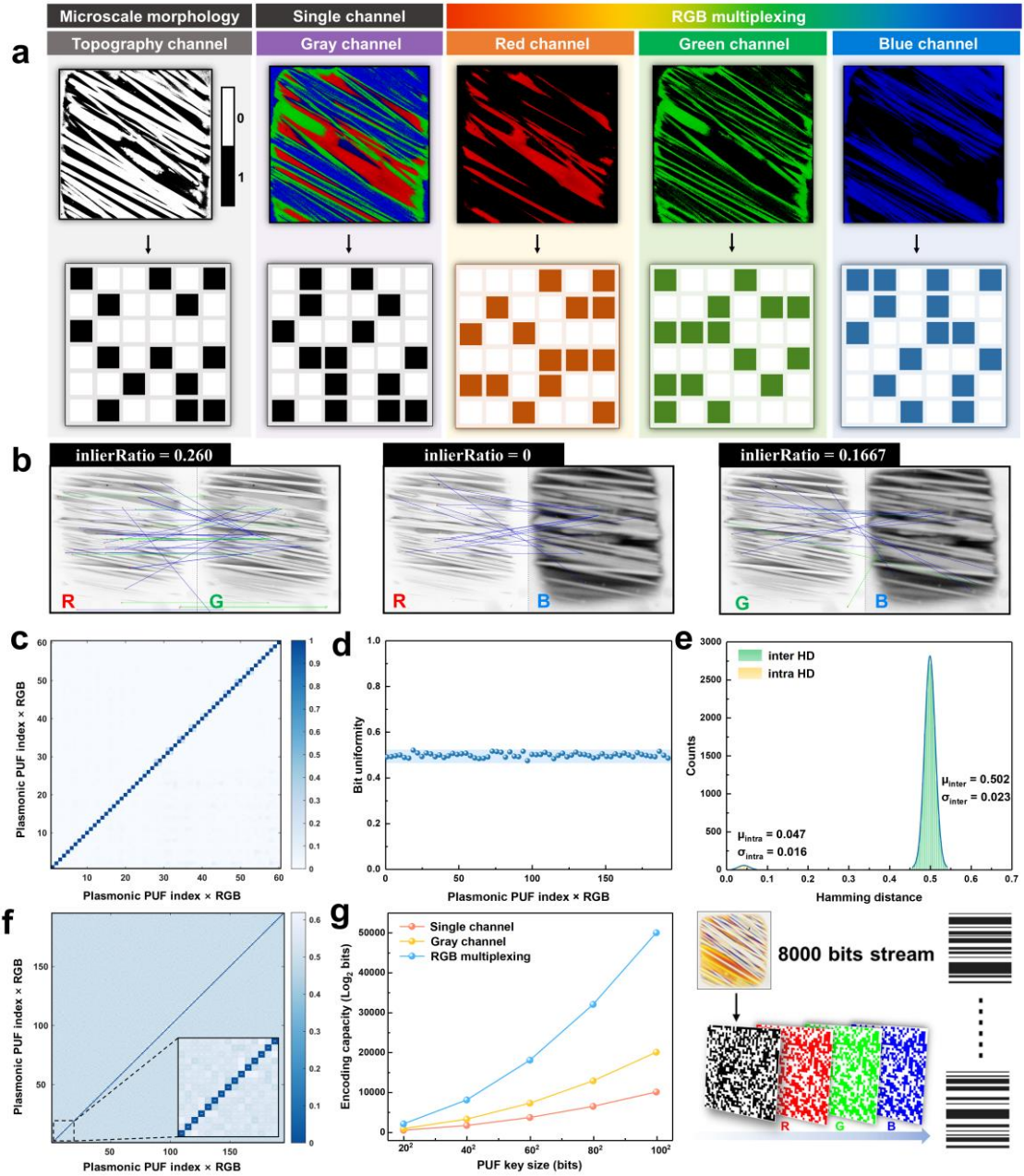

**Figure 4** │ RGB multiplexing plasmonic PUF keys. (a) Concept of multi-level key generation using one plasmonic PUF under signal challenge: from micro-scale morphology to single-channel grayscale and RGB multiplexed channels. (b) Feature matching by the SIFT algorithm for RGB channels of the same plasmonic PUF. (c) Heatmap of inlier ratios for SIFT feature matching between identical and different plasmonic PUF images across RGB channels. (d) Bit uniformity of all RGB channels of different plasmonic PUF keys. (e) inter- and intra-Hamming distance of the same and different PUF keys. (f) Similarity matrix of RGB multiplexed

plasmonic PUF keys using Hamming distance. (g) Left: Encoding capacity of one 40×40 bits PUF image by multimodal multiplexing, right: Schematic illustration of encoding capacity expanding strategy by splicing RGB channels.

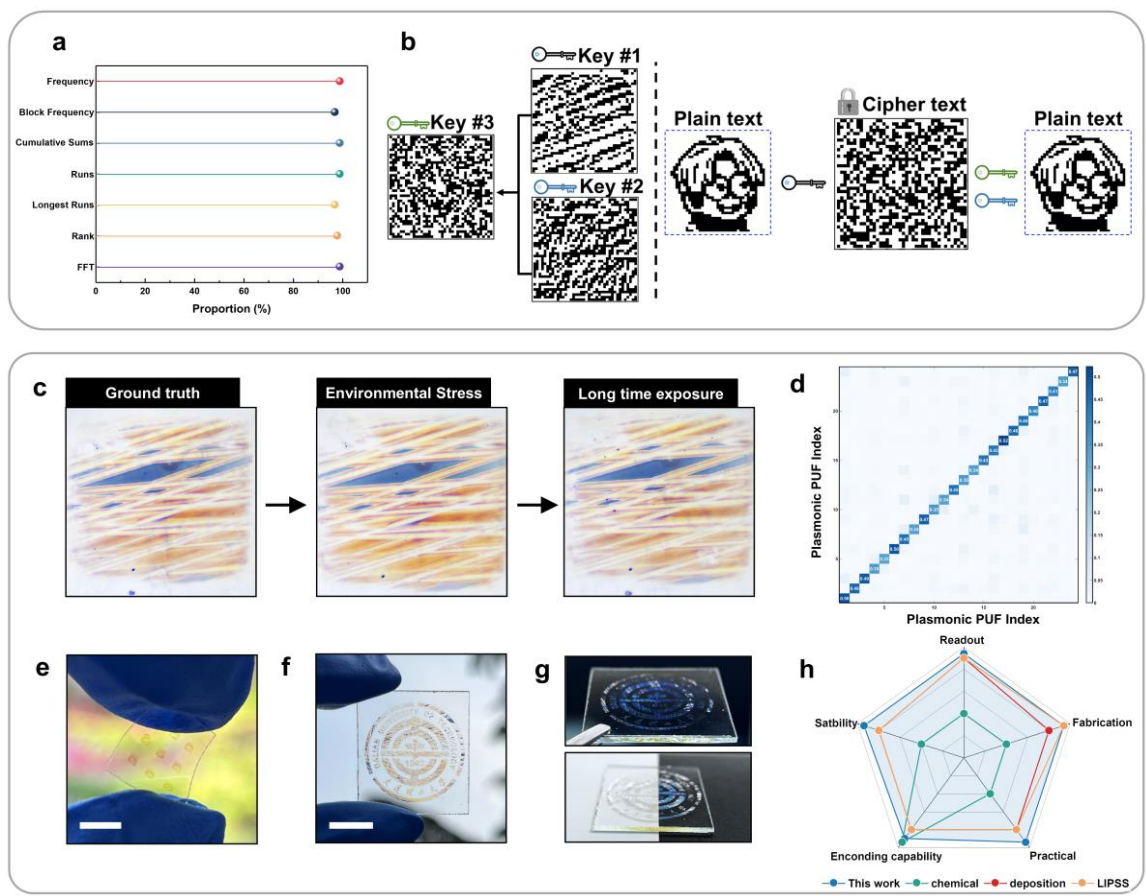

**Figure 5** | Practicality of plasmonic PUF keys. (a) Results of the NIST SP 800-22 statistical randomness test. (b) Schematic demonstration of a symmetric encryption application. (c) Optical image of the same plasmonic PUF before and after a series of extreme tests including high temperature, ultrasound vibration, and long-time light exposure. (d) Corresponding similarity matrix (inter-Hamming distances) calculated for the aged samples in (c), confirming that the authentication remains reliable (high diagonal correlation) even after rigorous durability testing. (e) Photographs showcasing the fabrication versatility on flexible substrates. Scale bar, 50 mm. (f) Scalability and macroscopic programmability demonstration featuring a large-area (25 mm × 25 mm) university logo. Scale bar, 250 mm. (g) Naked eye anti-counterfeiting: under oblique observation against a dark background, the patterned logo exhibits a vivid blue scattering color. (h) Radar chart comparing the comprehensive performance of this work (blue) with existing PUF strategies.